# Extracting grid characteristics from spatially distributed place cell inputs using non-negative PCA

*Yedidyah Dordek[1,2], Ron Meir [1], Dori Derdikman [2*]*

[1]*Faculty of Electrical Engineering, Technion - Israel Institute of Technology*

[2]*Rappaport Faculty of Medicine and Research Institute, Technion - Israel Institute of Technology*

*\* Correspondence should be addressed to Dori Derdikman, derdik@technion.ac.il*

## Abstract

Many recent models study the downstream projection from grid cells to place cells, while recent data has pointed out the importance of the feedback projection. We thus asked how grid cells are affected by the nature of the input from the place cells.

We propose a two-layered neural network with feedforward weights connecting place-like input cells to grid cell outputs. Place-to-grid weights were learned via a generalized Hebbian rule. The architecture of this network highly resembles neural networks used to perform Principal Component Analysis (PCA). Our results indicate that if the components of the feedforward neural network were non-negative, the output converged to a hexagonal lattice. Without the non-negativity constraint the output converged to a square lattice. Consistent with experiments, grid alignment to walls was ~7° and grid spacing ratio between consecutive modules was ~1.4. Our results express a possible linkage between place cell to grid cell interactions and PCA.

# Introduction

The system of spatial navigation in the brain has recently received much attention [1,2,3]. This system involves many regions, which seem to divide into two major classes: regions such as CA1 and CA3 of the hippocampus which contain place cells [4,5], vs. regions, such as the medial-entorhinal cortex (MEC), the presubiculum and the parasubiculum, which contain grid cells, head-direction cells and border cells [6,7,8,9,10]. While the phenomenology of those cells is described in many studies [11], the manner in which grid cells are formed is quite enigmatic. Many mechanisms have been proposed. The details of these mechanisms differ, however they mostly share in common the assumption that the animal's velocity is the main input to the system [11,12,13], such that positional information is generated by integration of this input in time. This process is termed "path integration" (PI)[14]. A notable exception to this class of models was suggested in a paper by Kropff & Treves [15], and in a sequel to that paper [16], in which they demonstrated the emergence of grid cells from place cell inputs without using the rat's velocity as an input signal.

We note here that generating grid cells from place cells may seem at odds with the architecture of the network, since it is known that place cells reside at least one synapse downstream of grid cells [17]. Nonetheless, there is current evidence that the feedback from place cells to grid cells is of great functional importance. Specifically, there is evidence that inactivation of place cells causes grid cells to disappear [18], and furthermore, it seems that, in development, place cells emerge before grid cells do [19,20]. Thus, there is good motivation for trying to understand how the feedback from hippocampal place cells may contribute to grid cell formation.

In the present paper we thus investigated a model of grid cell development from place cell inputs. We showed the resemblance between a feedforward network from place cells to grid cells to a neural network architecture previously used to implement the PCA algorithm [21]. We demonstrated that the formation of grid cells from place cells using such a neural network could occur given specific assumptions on the input (i.e. zero mean) and on the nature of the feedforward connections (specifically, non-negative, or excitatory).

# Results

## Comparing Neural-Network results to PCA

We initially considered the output of a two-layer neural network and of the PCA algorithm in response to the same inputs. These consisted of the temporal activity of a simulated agent moving around in a two-dimensional (2D) space (Fig 1A; see Methods for details). In order to mimic place cell activity, the simulated virtual space was covered by multiple 2D Gaussian functions uniformly distributed at random (Fig 1B), which constituted the input. In order to calculate the principal components we used a [Neuron x Time] matrix (Fig 1C) after subtracting the temporal mean, displaying a one-dimensional mapping of the two-dimensional activity transforming the 2D activity into a 1D vector per input neuron. This resulted in the [Neuron X Neuron] covariance matrix (Fig 1D), based on which PCA was performed by evaluating the appropriate eigenvalues and eigenvectors.

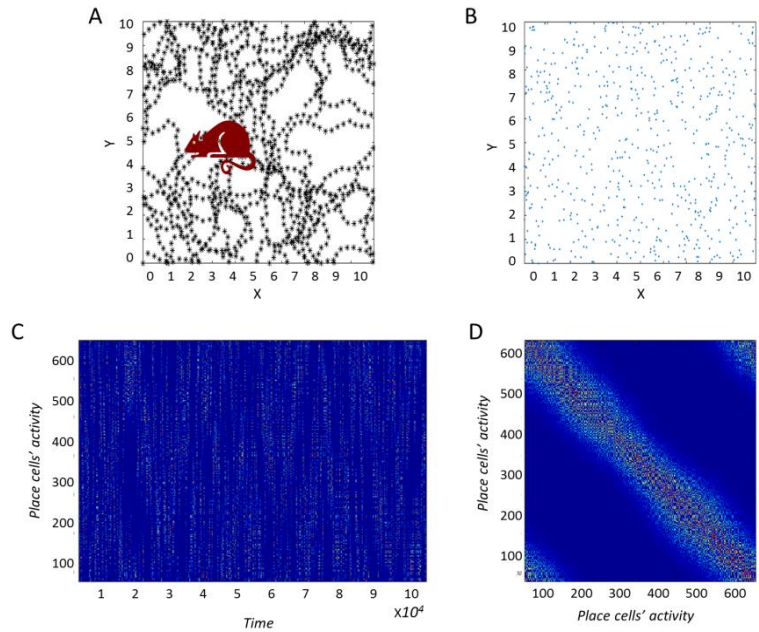

*Fig 1. Construction of the correlation matrix from behavior. (A) Diagram of the environment. Black dots indicate places the virtual agent has visited. (B) Centers of place cells uniformly distributed in the environment. (C) The [Neuron X Time] matrix of the input-place cells. (D) Correlation matrix of (C) used for the PCA process.*

To learn the grid cells based on the place cell inputs, we implemented a two-layered neural network with a single output (Fig 2). Input to output weights were governed by a Hebbian-like learning rule. As described in the Introduction (elaborated upon in the Methods section), this type of architecture induces the output's weights to converge to the leading principal component of the input data.

The agent explored the environment for a sufficiently long time allowing the weights to converge to the first principal component of the temporal input data. In order to establish a spatial interpretation of the eigenvectors (from PCA) or the weights (from the converged network) we projected both the PCA eigenvectors and the network weights onto the place cells space, producing corresponding spatial activity maps. The leading eigenvectors of the PCA and the network's weights converged to square-like periodic spatial solutions (Fig 3A-B).

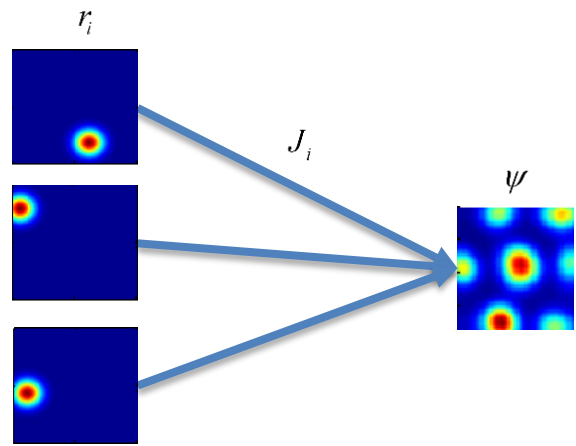

*Fig 2. Neural network architecture with feedforward connectivity. The input layer corresponds to place cells and the output layer to a single cell.*

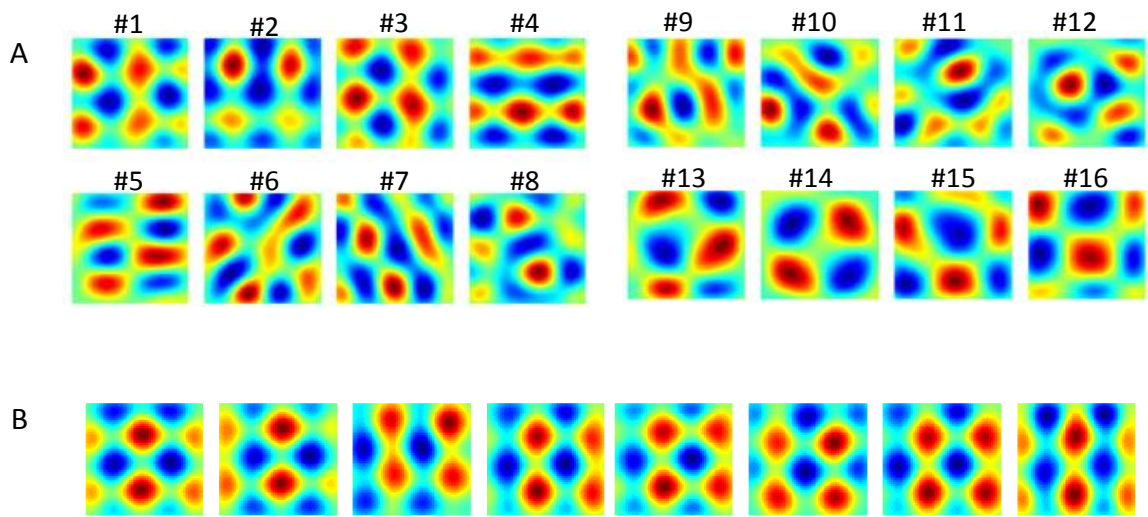

*Fig 3. Results of PCA and of the networks' output (in different simulations). (A) 1st 16 PCA eigenvectors projected on the place cells' input space. (B) Converged weights of the network (each result from different simulation, initial conditions and trajectory) projected onto place cells' space.*

Being a PCA algorithm, the spatial projections of the weights were periodic in space due to the covariance matrix of the input having a Toeplitz structure [22] (a Toeplitz matrix has constant elements along each diagonal). Intuitively, the Toeplitz structure arises due to the spatial stationarity of the input. In fact, since we used periodic boundary conditions for the agent's motion, the covariance matrix was a circulant matrix, and the eigenvectors were sinusoidal functions [23] (a circulant matrix is defined by a single row (or column), and the remaining rows (or columns) are obtained by cyclic permutations. It is a special case of a Toeplitz matrix). Due to the isotropic nature of the data (generated by the agent's motion), there was a 2D redundancy in the X-Y plane in conjunction with a dual phase redundancy in 1D, resulting in a 4-fold total redundancy of every solution. This was apparent in the plotted eigenvalues (from largest to the smallest eigenvalue, Fig 4A), which demonstrated a fourfold grouping-pattern. The covariance matrix was heavily degenerate with approximately 90% of the variance accounted for by the first 15% of the eigenvectors (Fig 4B).

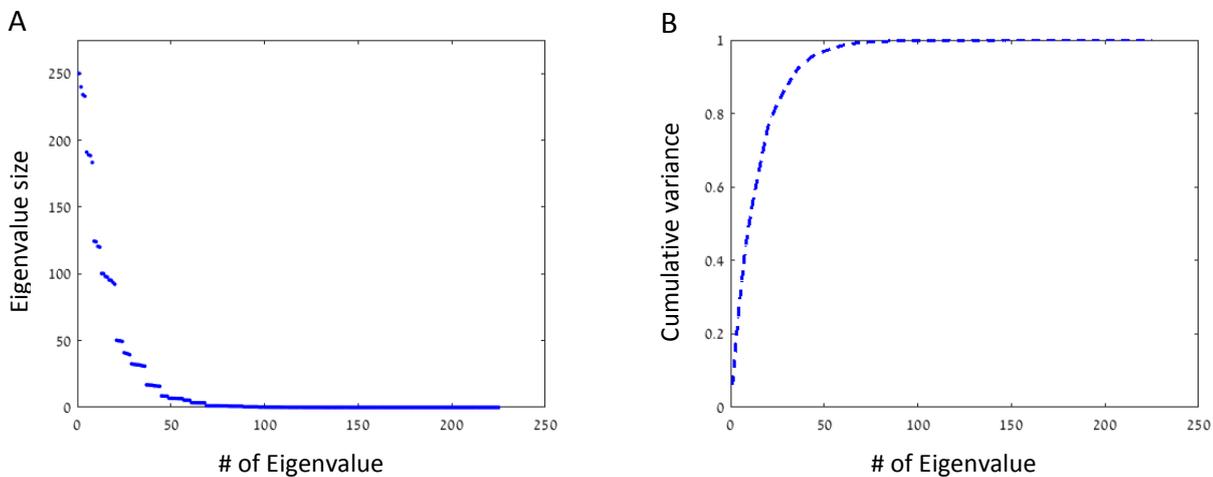

*Fig 4. Eigenvalues and eigenvectors of the input's correlation matrix. (A) 4-fold redundancy in the eigenvalues due to X-Y and phase degeneracy. (B) Cumulative explained variance by the eigenvalues, with 90% of variance accounted for by the first 15% eigenvectors. We used here only 225 inputs for improve readability of the 4-fold redundancy in (A).*

In summary, both the direct PCA algorithm and the neural network solutions developed periodic structure. However, this periodic structure was not hexagonal but rather had a square-like form.

## Adding a non-Negativity constraint to the PCA

It is known that most synapses from the hippocampus to the MEC are excitatory [17]. We thus checked how a non-negativity constraint, applied to the projections from place cells to grid cells, affected our simulations. We found that when adding this constraint, the outputs behaved in a different manner and converged to a **hexagonal grid**, similar to real grid cells. While it was straightforward to constrain the neural network, calculating non-negative PCA directly was an arduous task due to the non-convex nature of the problem [24,25].

In the network domain we used a simple cutoff rule for the learned feedforward weights, which constrained their values to be non-negative. For the direct non-negative PCA calculation, we used the raw place cells activity as inputs to two different iterative numerical methods: NSPCA (Nonnegative Sparse PCA) and an AMP (Approximate Message Passing) based algorithms (see Methods section).

In both cases we found that hexagonal grid cells emerged in the output layer (plotted as spatial projection of weights and eigenvectors: Fig 5A-B, Fig 6A-B). When we repeated the process over many simulations (i.e. new trajectories and random initializations of weights) we found that the population as a whole consistently converged to hexagonal grid-like responses, while similar simulations with the unconstrained version did not (compare Fig 3 to Fig 5 -Fig 6).

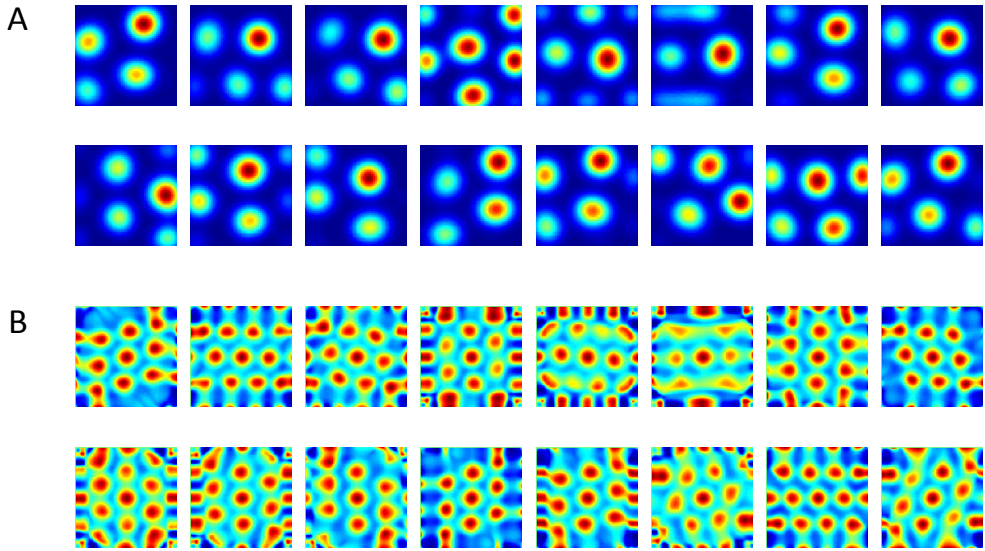

*Fig 5. Output of the neural network when weights are constrained to be non-negative. (A) Converged weights (from different simulations) of the network projected onto place cells space. (B) Spatial autocorrelations of (A).*

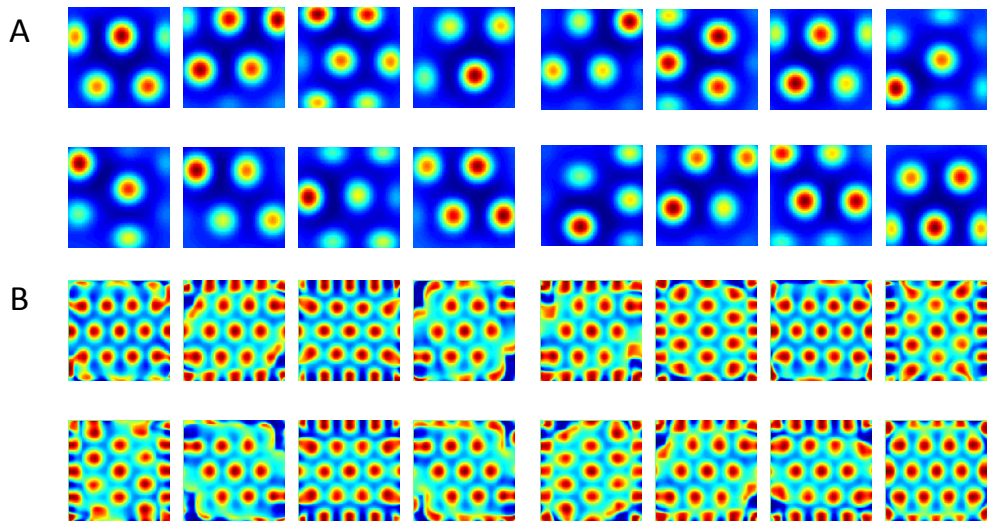

*Fig 6. Results from the non-negative PCA algorithm. (A) Spatial projection of the leading eigenvector on input space and the (B) corresponding spatial autocorrelations. The different solutions are outcomes of multiple simulations with identical settings in a new environment and new random initial conditions.*

In order to further assess the hexagonal grid emerging in the output, we calculated the mean (hexagonal) Gridness scores ([8], which measure the degree to which the solution resembles a hexagonal grid (see Methods)). We ran about 1500 simulations of the network (in each simulation, the network consisted of 625 place cell-like inputs and a single grid cell-like output), and found noticeable differences between the constrained and unconstrained cases. Namely, the Gridness score in the non-negatively constrained-weight simulations was significantly higher than in the unconstrained-weight case (Gridness = 1.01 ± 0.003 in the constrained case vs. 0.32 ± 0.002 in the unconstrained case. see Fig 7). A similar difference was observed with the direct PCA methods (1500 simulations, each with different trajectories, Gridness = 1.07 ± 0.0012 in the constrained case vs. 0.26 ± 0.0019 in the unconstrained case).

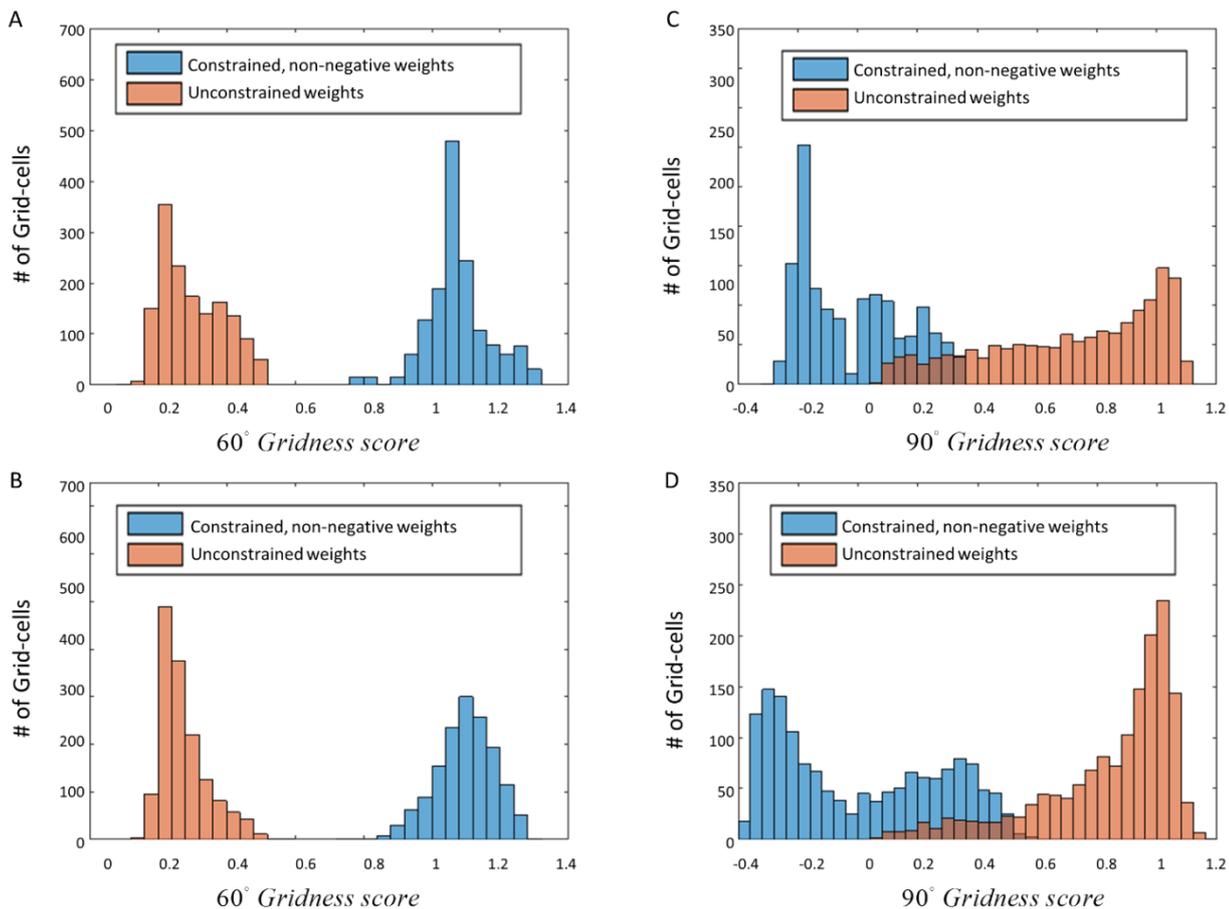

Fig 7. Histograms of Gridness values from network and PCA. First row (A) + (C) corresponds to network results, and second row (B) + (D) to PCA. The left column histograms contain the $60°$ Gridness scores and the right one the $90°$ Gridness scores.

Another score we tested was a "Square Gridness" score (see Methods) where we measured the "Squareness" of the solutions (as opposed to "Hexagonality"). We found that the unconstrained network had a higher square-Gridness score while the constrained network had a lower square-Gridness score (Fig 7); for both the direct-PCA calculation (square-Gridness = 0.81 ± 0.005 in the unconstrained case vs. 0.07 ± 0.005 in the constrained case) and the neural-network (square-Gridness = 0.073 ± 0.006 in the constrained case vs. 0.66 ± 0.007 in the unconstrained case).

All in all, these results suggest that when direct PCA eigenvectors and neural network weights were unconstrained they converged to periodic square solutions. However, when constrained to be non-negative, the direct PCA, and the corresponding neural network weights, both converged to a hexagonal solution.

## Dependence of the result on the structure of the input

We investigated the effect of different inputs on the emergence of the grid structure in the networks' output. We found that some manipulation of the input was necessary in order to enable the implementation of PCA in the neural network. Specifically, PCA requires a zero-mean input, while simple Gaussian-like place cells do not possess this property. In order to obtain input with zero-mean we either performed differentiation of the place cells' activity in time, or used a Mexican-hat like (Laplacian) shape (See Methods for more details on the different types of inputs). Another option we explored was the usage of positive-negative disks with a total sum of zero activity in space (Fig 8). The motivation for the use of Mexican-hat like transformations is their abundance in the nervous system [26,27,28].

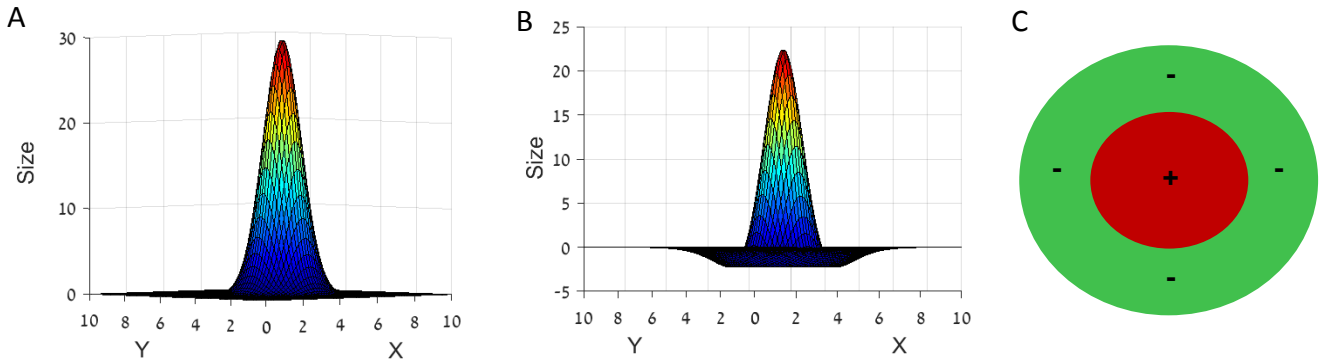

*Fig 8. Different types of input used in our network. (A) 2D Gaussian function, acting as a simple place cell. (B) Laplacian function or Mexican hat. (C) An inhibitory (outer ring) - excitatory (inner circle) disk.*

We found that usage of simple 2-D Gaussian-functions as inputs did not generate hexagonal grid cells as outputs (Fig 9). On the other hand, time-differentiated inputs, positive-negative disks or Laplacian inputs did generate grid-like output cells, both when running the non-negative PCA directly (Fig 6), or by simulating the non-negatively constrained Neural Network (Fig 5). Another approach we used for obtaining zero-mean was to subtract the mean dynamically from the every output individually (see Methods). The latter approach, related to adaptation of the firing rate, was adopted from Kropff & Treves [15], who used it in order to control various aspects of the grid cell's activity. In addition to controlling the firing rate of the grid cells, if applied correctly, the adaptation could be exploited to keep the output's activity stable, with zero-mean rates. We applied this method in our system and in this case the outputs converged to hexagonal grid cells as well, similarly to the previous cases (e.g. derivative in time, or Mexican hats as inputs; data not shown).

In summary, two conditions were required for the neural network to converge to spatial solutions resembling hexagonal grid cells: (1) non-negativity of the feedforward weights and (2) an effective zero-mean of the inputs (in time or space).

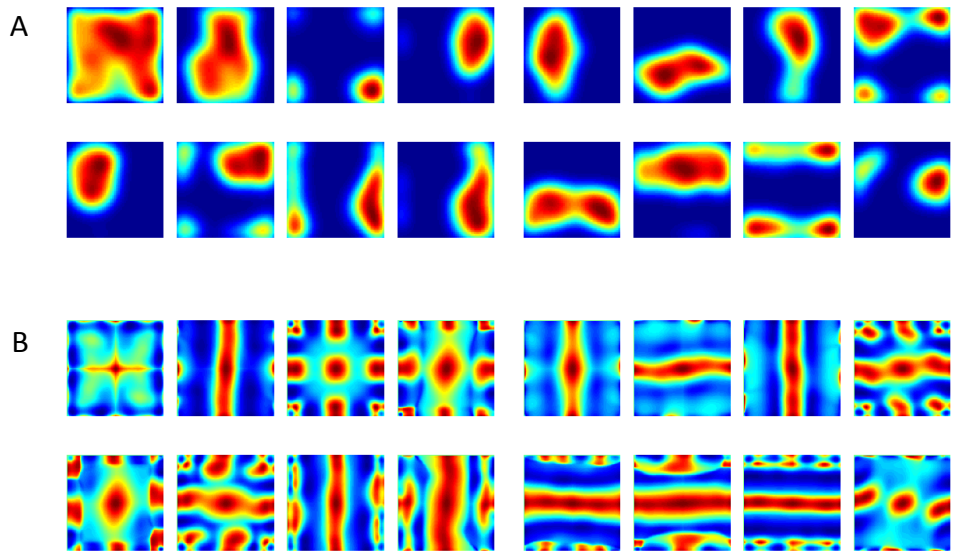

*Fig 9. Spatial projection of outputs' weights in the neural network when input did not have zero mean. (A) Various weights plotted spatially as projection onto place cells space. (B) Autocorrelation of (A).*

## Effect of place cell parameters on grid structure

A more detailed view of the resulting grid spacing showed that it was heavily dependent on the widths of the place cells inputs. When the environment size was fixed and the output calculated per input size, the grid-spacing (distance between neighboring peaks) increased for larger place cell widths. The outcome was consistent for all types of inputs, in particular DOGs (Difference of Gaussians; Fig 10B). The dependency was almost linear ($y = 6.5x + 0.16$). We further checked the angle of alignment of the grid cells to the walls. The latter was determined by the minimal degree between the three hexagon grid axes to one of the confining walls. While there was some fluctuation in the alignment of the grids to the wall, the overall mean alignment of the grid direction to the wall was 6.95 ± 1.31° (Fig 10A).

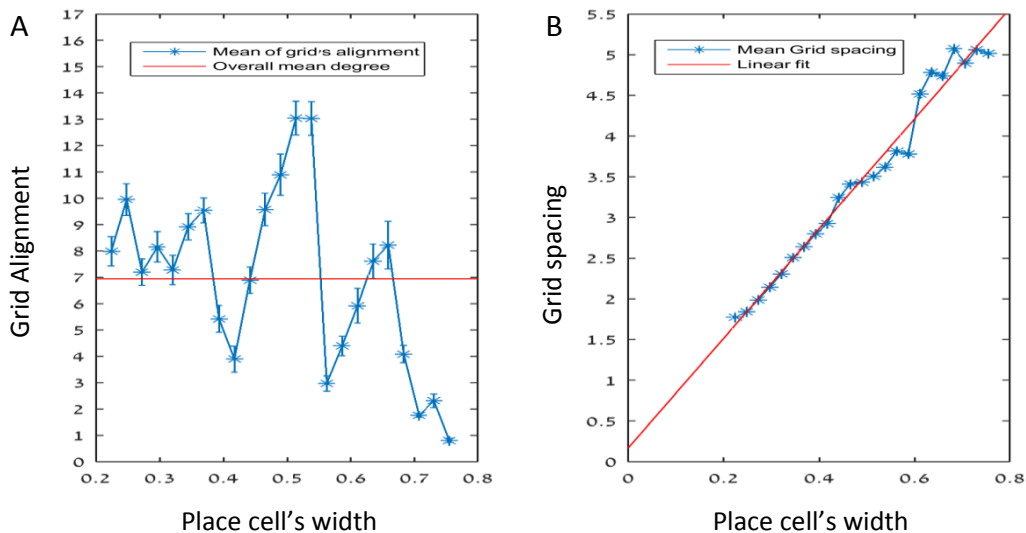

*Fig 10: Dependence of grid spacing and alignment on place cells' width. (A) Grid alginment fluctuates as place cells widhts vary (in blue). The overall mean of alignment degree was ~7° (in red) . (B) An almost linear dpendency between the place cells widths' and the grid cell's spacing. The X-axis in both panels was limited by resolution (lower bound) and ratio of place cell width to environment size (upper bound).*

## Modules of grid cells

It is known that in reality grid cells form in modules of multiple spacings [29,30]. We tried to address this question of modules (i.e., multiple grid spacings when assuming various settings) in several ways. First, we used different widths for the Gaussian/Laplacian input functions: Initially, we placed a heterogeneous population of widths in a given environment (i.e., uniformly random widths) and ran the single-output network 100 times. The distribution of grid spacings was almost comparable to the results of the largest width if applied alone, and did not exhibit module like behavior. This result is not surprising when thinking about a small place cell overlapping in space with a large place cell. Whenever the agent passes next to the small one it activates both weights via synaptic learning. This causes the large firing field to overshadow the smaller one. Additionally, when using populations of only two widths of place fields, the grid spacings were dictated by the size of the larger place field (data not shown).

The second option we considered was to use a multi-output neural network, capable of computing all "eigenvectors" rather than only the principal "eigenvector" (where by "eigenvector" we mean here the vectors achieved under the positivity constraint, and not the exact eigenvectors themselves). We used a hierarchical network implementation introduced by [35] (see Methods). Since the 1st outputs' weights converged to the 1st "eigenvector", the network (Fig 11 A-B) provided to the subsequent outputs (2nd, 3rd, and so forth) a reduced-version of the data from which the projection of the 1st "eigenvector" has been subtracted out. This process, reminiscent of Gram-Schmidt orthogonalization, was capable of computing all "eigenvectors" (in the modified sense) of the input's covariance matrix.

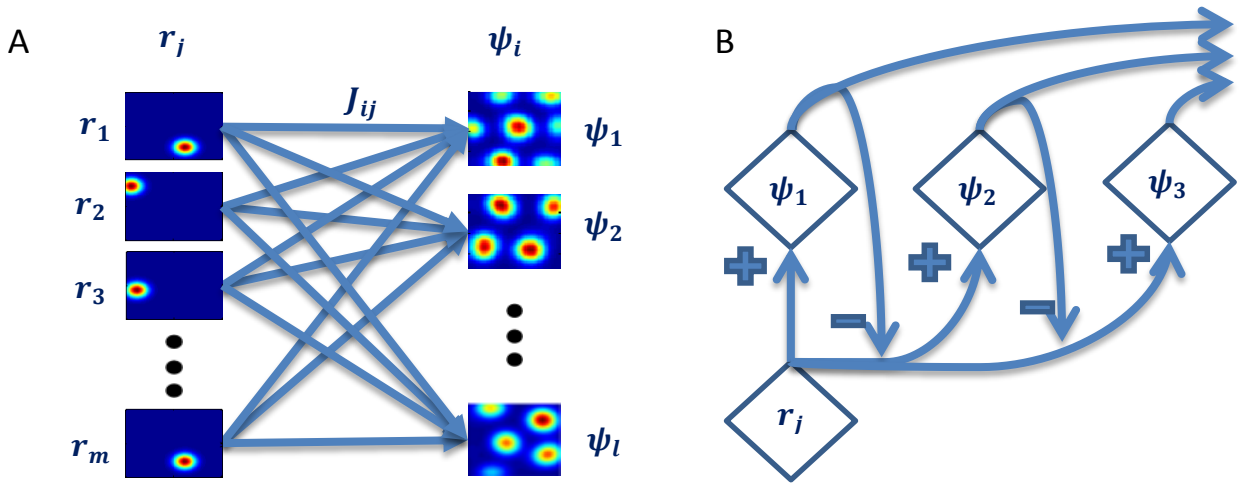

*Fig 11. Hierarchial netwrok capable of computing all "principal components". (A) Each output is a linear sum of all inputs weighted by the corresponding learned weights. (B) Over time, the data the following outputs "see" is the original data after subtration of the 1st "eigenvector's" projection onto it. This is an iterative process causing all outputs' weights to converge to the "prinipcal components" of the data.*

When constrained to be non-negative, and using the same homogeneous "place cells" as in the previous network, the networks' weights converged to hexagonal shapes. Here however, we found that the smaller the "eigenvalue" was (or the higher the principal component number) the denser the grid became. We were able to identify two main populations of grid-distance "modules" among the hexagonal spatial solutions with high Gridness scores (above 0.7, Fig 12A-B). In addition, we found that the ratio between the distances of the modules was ~1.4, close to the value of 1.42 found by Stensola et al. [30]. The same process was repeated using the direct PCA method, utilizing the covariance matrix of the data after simulation as input for the non-negative PCA algorithms, and considering their ability to calculate only the $1^{st}$ "eigenvector". By iteratively projecting the $1^{st}$ "eigenvector" on the simulation data and subtracting the outcome from the original data, we applied the non-negative PCA algorithm to the residual data obtaining the $2^{nd}$ "eigenvector" of the original data. This "eigenvector" now constituted the $1^{st}$ "eigenvector" of the new residual data (see Methods). Applying this process to as many "outputs" as needed, we obtained very similar results to the ones presented above using the neural network (data not shown).

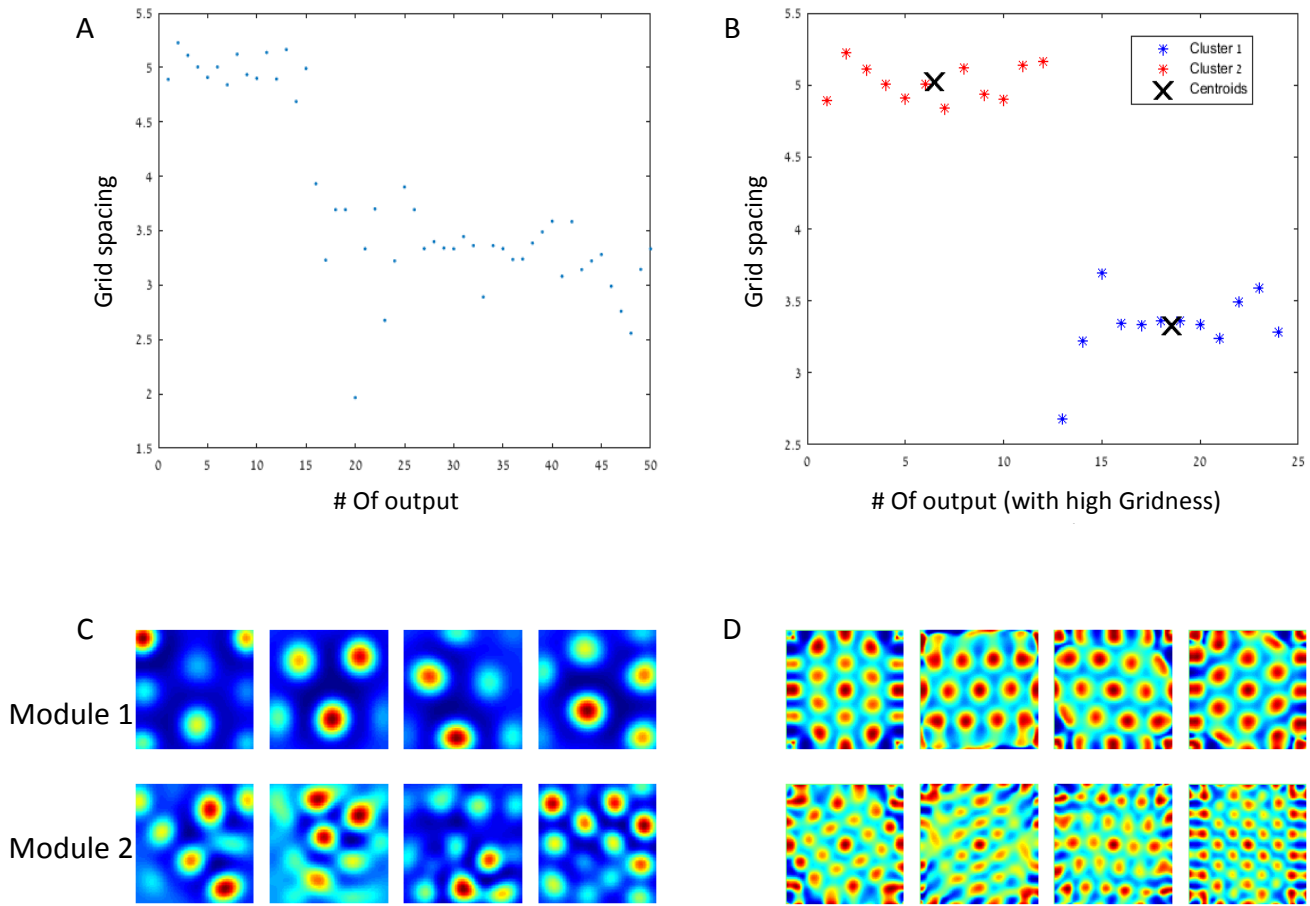

*Fig 12. Modules of grid cells. (A) In a network with 50 outputs, the grid spacing per output is plotted with respect to the hiererchial place of the output. (B) The grid spacing of outputs with high Gridness score (>0.7). The centroids have a ratio of close to √2. (C) + (D) Example of rate maps of outputs and their spatial autocorrelations for both of the modules.*

## Stability analysis

### Convergence to hexagons from various initial spatial conditions

In order to numerically test the stability of the hexagonal solution, we initialized the network in different ways, randomly, using linear stripes, squares, rhomboids (squares on hexagonal lattice) and noisy hexagons. In all cases the network converged to a hexagonal pattern (Fig 13; for squares and strips, other shapes not shown here).

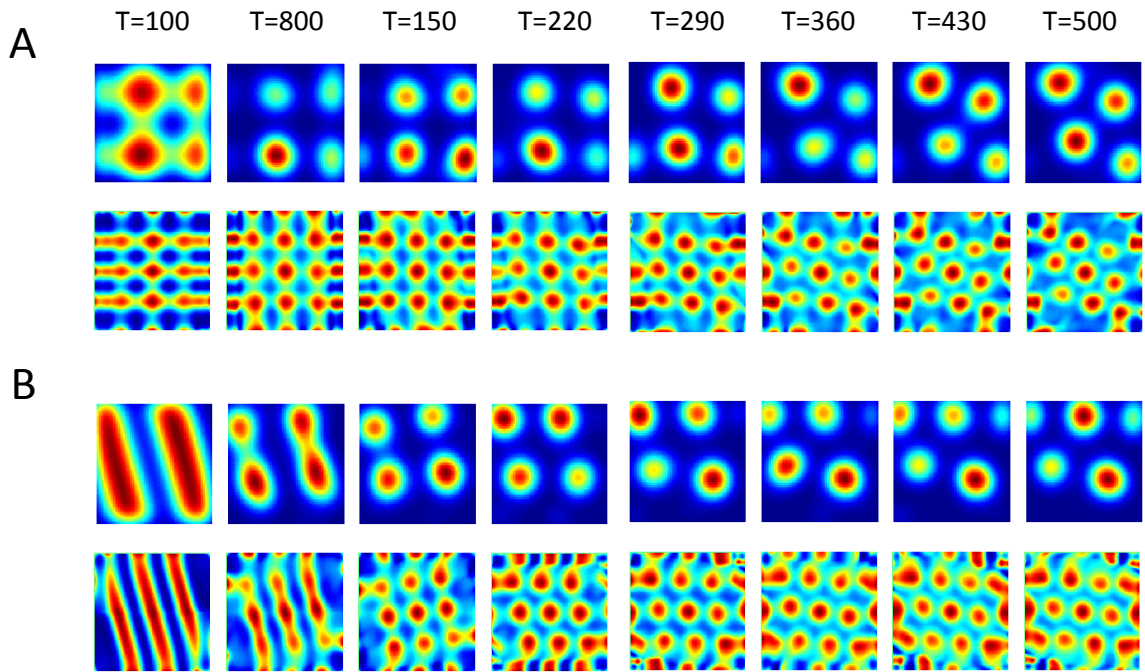

*Fig 13: Evolution in time of the networks' solutions (upper rows) and their autocorrelations (lower rows). The network was initialized in shapes of (A) Squares and of (B) stripes (linear).*

We also ran the converged weights in a new simulation with novel trajectories and tested the Gridness scores, and the inter-trial stability in comparison to previous simulations. We found that the hexagonal solutions of the network remained stable although the trajectories varied drastically (data not shown).

## Asymptotic stability of the equilibria

Under certain conditions (e.g., decaying learning rates and independent and identically distributed (i.i.d.) inputs) it was previously proved [31], using techniques from the theory of stochastic approximation, that the system described here can be asymptotically analyzed in terms of (deterministic) Ordinary Differential Equations (ODE), rather than in terms of the stochastic recurrence equations. Since the ODE defining the converged weights is non-linear, an analytical characterization was difficult to achieve. Instead, we solved the ODEs numerically (see Methods), by randomly initializing the weight vector. The asymptotic equilibria were reached much faster, compared to the outcome of the recurrence equations. Similarly to the recurrence equations, constraining the weights to be non-negative induced them to converge into a hexagonal shape while a non-constrained system produced square-like outcomes (Fig 14).

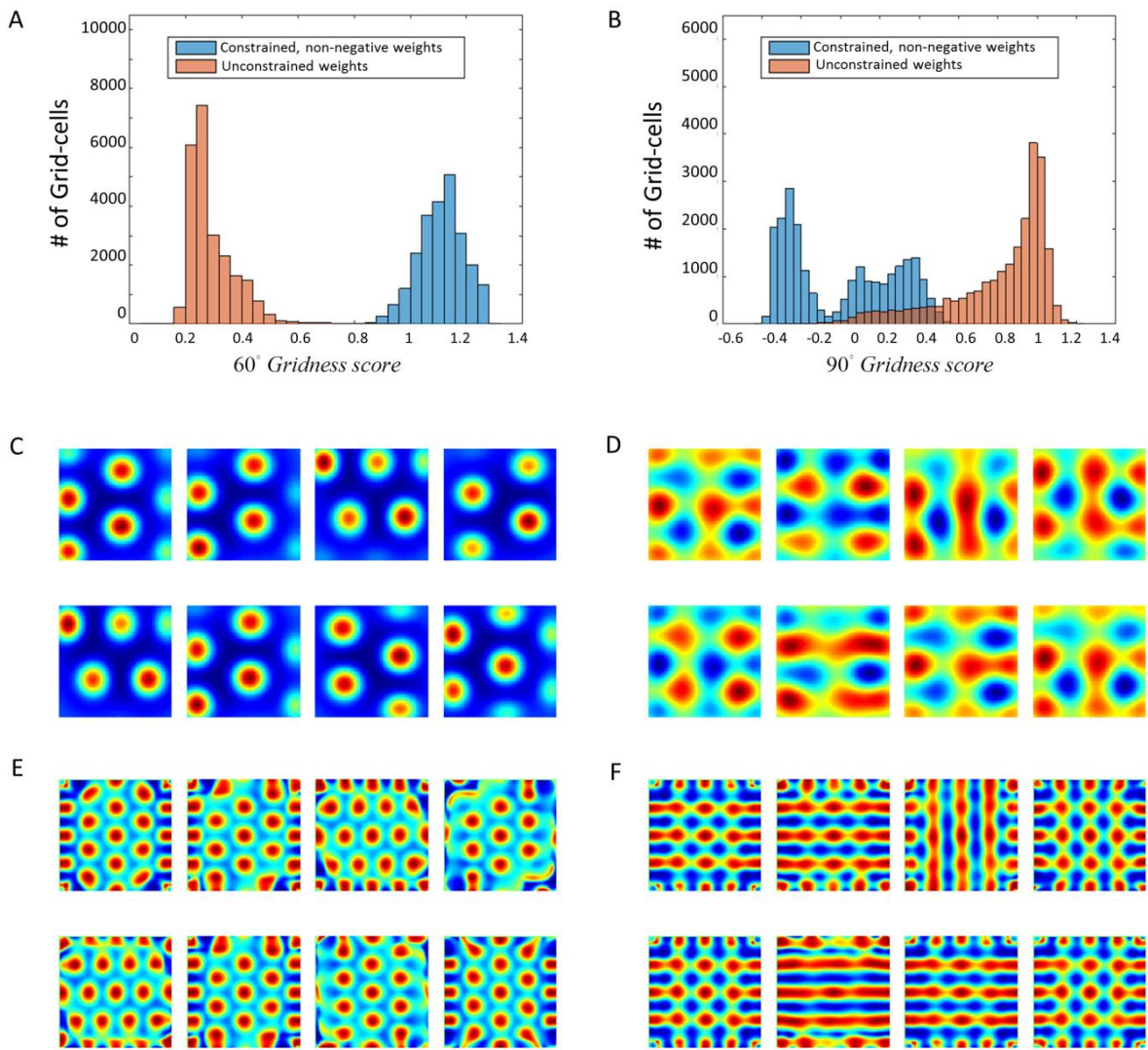

*Fig 14. Numerical convergence of the ODE to hexagonal results when weights are constrained. (A) + (B): 60° and 90° Gridness score histograms. Each score represents a different weight vector of the solution J. (C) + (D): Spatial results for constrained and unconstrained scenarios, respectively. (E) + (F) Spatial autocorrelations of (C) + (D).*

Simulation was run 60 times, with 400 outputs per run. $60°$ Gridness score mean was 1.1 ± 0.0006 when weights were constrained and 0.29 ± 0.0005 when weights were unconstrained. $90°$ Gridness score mean was 0.006 ± 0.002 when weights were constrained and 0.8 ± 0.0017 when weights were unconstrained.

# Discussion

In our work we explored the nature and behavior of the feedback projections from place cells to grid cells. We shed light on the importance of this relation and showed how a simple two-layered neural network could produce hexagonal grid cells when subjected to place cell-like temporal input from a randomly-walking moving agent.

## Place-to-Grid as a PCA network

We noticed that such a neural network was capable of finding the Principal Components of the input data [21]. We thus asked, whether the interaction of grid cells and place cells could be interpreted as a process of performing PCA.

As a consequence of the requirements for PCA to hold, we found that the place cell input needed to possess the quality of zero-mean, otherwise the output was not periodic. Due to the lack of the zero-mean property in 2D Gaussians we used various approaches to impose zero-mean on the input data. The first, in the time domain, was to differentiate the input and use the derivatives (a random walk produces zero-mean derivatives) as inputs. Another approach was to dynamically subtract the mean in all iterations of the simulation. This approach was reminiscent of the adaptation procedure suggested in the Kropff & Treves paper [15]. A third approach, applied in the spatial domain was to use inputs with a zero-spatial mean such as Laplacians of Gaussians (Mexican hats in 2D, or differences-of-Gaussians) or negative – positive disks. Such Mexican-hat inputs are quite typical in the nervous system [26,27,28], although in the case of place cells it is not completely known how they are formed. They could be a result of interaction between place cells and the vast number of inhibitory interneurons in the local hippocampal network [32].

Another condition we found crucial, which was not part of the original PCA network, was a non-negativity constraint on the place-to-grid learned weights. While rather easy to implement in the network, adding this constraint to the non-convex PCA problem was harder to implement. Since the problem is NP-hard [24] we turned to numerical methods. We used two different algorithms [24,33] to find the leading "eigenvector" of every given temporal based input. As shown in the results section, both processes (i.e. direct PCA and the neural network) resulted in hexagonal outcomes when the non-negativity and zero-mean criteria were met.

We note that while our network focused on the projection from place cells to grid cells, we cannot preclude the importance of the reciprocal projection from grid cells to place cells. Further study will be needed in order to 'close the loop' and simultaneously consider both of these projections at once.

## Predictions of our model

Based on these findings, it is possible to make several predictions. First, the grid cells must receive zero-mean input over time in order to produce hexagonally shaped firing patterns. With all feedback projections from place cells being excitatory, the lateral inhibition from other neighboring grid cells might be the balancing parameter to achieve the temporal zero-mean. Alternatively, an adaptation method, such as the one suggested in Kropff & Treves [15] may be applied. Second, if indeed the grid cells are a lower dimensional representation of the place cells in a PCA form, the place-to-grid neural weights distribution should be similar across identically spaced grid cell population. This is because all grid cells with similar spacing would have maximized the variance over the same input, resulting in similar spatial solutions. Third, we found a constant relation between the size of the place cells and the spacing between grid cells. Furthermore, the spacing of the grid cells is mostly determined by the size of the largest place cell – predicting that the feedback from large place cells is not connected to grid cells with small spacing. Fourth, we found that the mean alignment of the grids with the walls was about 7° (Fig 10), which closely resembles the alignment in real grid cells [34]. Fifth, we found modules of different grid spacings in a hierarchical network with the ratio of distances between successive units close to $\sqrt{2}$. This result is in accord with the ratio reported in Stensola et al. [30].

## Why Hexagons?

In light of our results, we further asked what is special about the hexagonal shape which renders it a stable solution. Past works have demonstrated that hexagonality is optimal in terms of efficient coding. Two recent papers have addressed the potential benefit of encoding by grid cells. Mathis et al. [35] considered the decoding of spatial information based on a grid-like periodic representation. Using lower bounds on the reconstruction error based on a Fisher information criterion, they demonstrated that hexagonal grids lead to the highest spatial resolution in two dimensions (extensions to higher dimensions were also provided). The solution is obtained by mapping the problem onto a circle packing problem. The work of Wei et al. [36] also took a decoding perspective, and showed that hexagonal grids minimize the number of neurons required to encode location with a given resolution. Both papers offer insights into the possible information theoretic benefits of the hexagonal grid solution. In the present paper we were mainly concerned with a specific biologically motivated learning (development) mechanism that may yield such a solution. We demonstrated that hexagonality may originate from a maximum-variance principle, in which the hexagonal output is optimal in the sense of optimizing variability.

To conclude, this work demonstrates how grid cells could be formed from a simple Hebbian neural network with place cells as inputs, without needing to rely on path-integration mechanisms.

## Methods

### Neural Network Architecture

We implemented a two-layered neural network with feedforward connections that was capable of producing a hexagonal-like output (Fig 2). The feedforward connections were updated according to a self-normalizing version of a Hebbian learning rule referred to as the Oja rule [21],

$$\Delta J_i^t = \varepsilon^t \left( \psi^t r_i^t - \psi^t \cdot J_i^t \right) \qquad (1)$$

where $\varepsilon^t$ denotes the learning rate, $J_i^t$ is the $i^{th}$ weight and $\psi^t, r_i^t$ are the output and the $i^{th}$ input of the network, respectively (all at time $t$). The output $\psi^t$ was calculated every iteration by summing up all pre-synaptic activity from the entire input neuron population. The activity of each output was processed through a sigmoidal function (e.g., $\tanh$ ) or a simple linear function. Formally,

$$\psi^t = f\left( \sum_{i=1}^{n} J_i^t \cdot r_i^t \right). \qquad (2)$$

Since we were initially only concerned with the eigenvector associated with the largest eigenvalue, we did not implement a multiple-output architecture. When no lateral weights were used, multiple outputs were equivalent to running the same setting with one output several times.

As discussed in the introduction, this kind of simple feedforward neural network with linear activation is known to perform *Principal Components Analysis (PCA)* [21,37,38]. In other words, the feedforward weights converge to the eigenvector of the input's covariance matrix (or, in certain cases [38], to the subspace spanned by the principal eigenvectors). We can thus compare the results of the neural network to those of the mathematical procedure of PCA. Hence, in our simulation, we (1) let the neural networks' weights develop in real time based on the current place cell inputs. In addition, we (2) saved the input activity for every time step to calculate the input covariance matrix and perform (batch) PCA directly. The question we therefore asked was under what conditions, when using *place cell-like* inputs, a solution resembling hexagonal *grid cells* emerges. To answer this we used both the neural-network implementation and the direct calculation of the PCA coefficients.

## Simulation

We simulated an agent moving in a 2D virtual environment consisting of a square arena covered by $n$ uniformly distributed 2D Gaussian-shaped place cells given by

$$r_i^t(\underline{X}(t)) = \exp\left(\frac{-(\underline{X}(t)-\underline{C_i})^2}{2\sigma_i^2}\right), \quad i=1,2,\ldots,n \qquad (3)$$

where $X(t)$ represents the location of the agent. The variables $r_i^t$ constitute the temporal input from place cell $i$ at time $t$, and $C_i, \sigma_i$ are the $i^{th}$ place cell's center and width, respectively (see variations on this input structure below). In order to eliminate boundary effects, periodic boundary conditions were assumed. The virtual agent moved about in a random walk scheme (see Appendix) and explored the environment (Fig 1A). The place cell centers were assumed to be uniformly distributed (Fig 1B) and shared the same standard deviation $\sigma$. The activity of all place cells as a function of time $(r(t)_1, r(t)_2 \ldots r(t)_n)$ was dependent on the stochastic movement of the agent, and formed a [*Neuron x Time*] matrix ($r \in R^{nxT}$ $T$- being the Time dimension, see Fig 1C).

The simulation was run several times with different input arguments (see table 1). The agent was simulated for $T$ time steps, allowing the neural network's weights to develop and reach a steady state by using the learning rule (equations (1 - 2)) and the input (equation (3)) data. The simulation parameters are listed below and include parameters related to the environment, simulation, agent and network variables.

| **Environment:** | Size of arena | Place cells width | Place cells distribution |
|---|---|---|---|
| **Agent:** | Velocity (angular & linear) | Initial position | ------------------- |
| **Network:** | # Place cells/ #Grid cells | Learning rate | Adaptation variable (if used) |
| **Simulation:** | Duration (time) | Time step | ------------------- |

Table 1: List of variables used in simulation

To calculate the PCA directly, we used the MATLAB function **Princomp** in order to evaluate the $n$ principal eigenvectors $\{\vec{q}_k\}_{k=1}^n$ and corresponding eigenvalues of the input covariance matrix. As mentioned in the Results section there exist a near four-fold redundancy in the eigenvectors (X-Y axis and in phase). Fig 3 demonstrates this redundancy by plotting the eigenvalues of the covariance matrix. The output response of each eigenvector $\vec{q}_k$ corresponding to a 2D input location $(x, y)$ is

$$\Phi(x,y)_k = \sum_{j=1}^{n} q_k^j \exp\left(-\frac{(x-c_x^j)^2}{2\sigma_x^2} - \frac{(y-c_y^j)^2}{2\sigma_y^2}\right), \quad k=1,2,...,n \qquad (4)$$

## Non-negativity constraint

Projections between place cells and grid cells are known to be primarily excitatory [17], thus if we aim to mimic the biological circuit, a non-negativity constraint should be added to the feedforward weights in the neural network. While implementing a non-negativity constraint in the neural network is rather easy (e.g., a simple cutoff rule in the weight dynamics), the equivalent condition for calculating non-negative Principal Components is more intricate. Since this problem is non-convex and, in general, NP-hard [24], a numerical procedure was imperative. We used two different algorithms for this purpose.

The first [33] named NSPCA (Nonnegative Sparse PCA) is based on coordinate-descent. The algorithm computes a non-negative version of the covariance matrix's eigenvectors and relies on solving a numerical optimization problem, converging to a local maximum starting from a random initial point. The local nature of the algorithm did not guarantee a convergence to a global optimum (recall that the problem is non-convex). The algorithm's inputs consisted of the place cell activities' covariance matrix, $\alpha$ - a balancing parameter between reconstruction and orthonormality, $\beta$ – a variable which controls the amount of sparseness required, and an initial solution vector. For the sake of generality, we set the initial vector to be uniformly random (and normalized), $\alpha$ was set to a relatively high value – $10^4$ and since no sparseness was needed, $\beta$ was set to zero.

The second algorithm [24] does not require any simulation parameters except an arbitrary initialization. It works directly on the inputs and uses a message passing algorithm to define an iterative algorithm to approximately solve the optimization problem. Under specific assumptions it can be shown that the algorithm asymptotically solves the problem (for large input dimensions).

## Different variants of input structure

Performing PCA on raw data requires the subtraction of the data mean. Some thought was required in order to determine how to perform this subtraction in the case of the neural network.

One way to perform the subtraction in the time domain was to dynamically subtract the mean during simulation by using the discrete 1st or 2nd derivatives of the inputs in time (i.e. from equation (3) $\Delta r(t+1) = r(t+1) - r(t)$). Under conditions of an isotropic random walk (namely, given any starting position, motion in all directions is equally likely) it is clear that $E[\Delta r(t)] = 0$. Another option for subtracting the mean in the time domain was the use of an adaptation variable, as was initially introduced by [15]. Although originally exploited for control over the firing rate, it can be viewed as a variable that represents subtraction of a weighted sum of the firing rate history. Instead of using the inputs $r_i^t$ directly in equation (3) to compute the activation $\psi^t$, an intermediate adaptation variable $\psi_{adp}^t(\delta)$ was used ($\delta$ being the relative significance of the present temporal sample) as

$$\psi_{adp}^t = \psi^t - \bar{\psi}^t \tag{5}$$

$$\bar{\psi}^t = (1-\delta) \cdot \bar{\psi}^{t-1} + \delta \bar{\psi}^t \tag{6}$$

It is not hard to see that for *i.i.d.* variables $\psi_{adp}^t$, the sequence $\bar{\psi}_i^t$ converges for large $t$ to the mean of $\psi^t$. Thus, when $t \to \infty$ we find that $E[\psi_{adp,i}^t] \to 0$, specifically, the adaptation variable is of zero asymptotic mean.

The second method we used to enforce a zero mean input was simply to create it in advance. Rather than using 2D Gaussian functions (i.e. equation (3)) as inputs we used 2D difference-of-Gaussians (all $\sigma$ are equal in x and y axis):

$$r_i^t(\underline{X}(t)) = \exp\left(-\frac{(\underline{X}(t)-\underline{C}_i)^2}{2\sigma_{1,i}^2}\right) - \frac{\sigma_{1,i}^2}{\sigma_{2,i}^2} \cdot \exp\left(-\frac{(\underline{X}(t)-\underline{C}_i)^2}{2\sigma_{2,i}^2}\right), \quad i=1,2,\dots,n \tag{7}$$

It is easy to notice that the integral of the given Laplacian function is zero and if we assume a random walk that covers the entire environment uniformly, the temporal mean of the input would be zero as well. Such input data can be inspired by similar behavior of neurons in the retina and the lateral-geniculate nucleus [26,27]. Finally, we implemented another input data type; positive-negative disks (see Appendix). Analogously to the Difference-of-Gaussian functions, the integral over input is zero so the same goal (zero-mean) was achieved. It is worthwhile noting that subtracting a constant from a simple Gaussian function is not sufficient since at infinity it does not reach zero.

## Quality of solution and Gridness

In order to test the hexagonality of the results we used a hexagonal *Gridness score* [8]. The Gridness score of the spatial fields was calculated from a cropped ring of their autocorrelogram including the six maxima closest to the center. The ring was rotated six times, 30° per rotation, reaching in total angles of 30°, 60°, 90°, 120°, 150°. Furthermore, for every rotated angle the Pearson correlation with the original un-rotated map was obtained. Denoting by $C_\gamma$ the correlation for a specific rotation angle $\gamma$, the final Gridness score was [15]:

$$Gridness = \frac{1}{2}(C_{60} + C_{120}) - \frac{1}{3}(C_{30} + C_{90} + C_{150}) \tag{8}$$

In addition to this "traditional" score we used a *Squareness* Gridness score in order to examine how square-like the results are spatially. The special reference to the square shape was driven by the tendency of the spatial solution to converge to a rectangular shape when no constrains were applied. The Squareness Gridness score is similar to the hexagonal one, but now the cropped ring of the autocorrelogram is rotated 45° every iteration to reach angles of 45°, 90°, 135°. As before, denoting $C_\gamma$ as the correlation for a specific rotation angle $\gamma$ the new Gridness score was calculated as:

$$Square\,Gridness = C_{90} - \frac{1}{2}(C_{45} + C_{135}) \qquad (9)$$

All errors calculated in gridness measures are SEM (Standard Error of the Mean).

## Hierarchical networks and modules

As described in the Results section, we were interested to check whether a hierarchy of outputs could explain the module phenomenon described for real grid cells. We replaced the single-output network with a hierarchical, multiple outputs network, which is capable of computing all "principal components" of the input data while maintaining the non-negativity constraint as before. The network, introduced by [35], computes each output as a linear summation of the weighted inputs similar to eq. (2). However, the weights are now calculated according to:

$$\Delta J_{ij}^t = \varepsilon^t \left( r_j^t \psi_i^t - \psi_i^t \sum_{k=1}^{i} J_{kj}^t \psi_k^t \right) \qquad (10)$$

The first term in the parenthesis when $k = 1$ was the regular Hebb-Oja derived rule. In other words, the first output calculated the first non-negative "principal component" (in inverted commas due to the non-negativity) of the data. Following the first one, the weights of each output received a back projection from the previous outputs. This learning rule applied to the data in a similar manner to the Gram-Schmidt process, subtracting the "influence" of the previous "principal components" on the data and recalculating the appropriate "principal components" of the updated input data.

In a comparable manner, we applied this technique to the input data $X$ in order to obtain non-negative "eigenvectors" from the direct nonnegative-PCA algorithms. Bearing in mind that the second "eigenvector" $V_2$ was perpendicular to the first one $V_1$ and that the algorithms are capable of calculating only the first principal component of a given data, we found $V_2$ by subtracting from the data the projection of $V_1$ on it,

$$\tilde{X} = X - V_1^T (V_1 \cdot X) .  \qquad (11)$$

Next, we computed $V_2$, the first non-negative "principal component" of $\tilde{X}$, and similarly the subsequent ones.

## Stability of Hexagonal solutions

In order to test the stability of the solutions we obtained under all types of conditions, we applied the ODE method [25,31,38] to the PCA feature extraction algorithm introduced in pervious sections. This method allows one to asymptotically replace the stochastic update equations describing the neural dynamics by smooth differential equations describing the average asymptotic behavior. Under appropriate conditions, the stochastic dynamics converge with probability one to the solution of the ODEs. Although originally this approach was designed for a more general architecture (including lateral connections and asymmetric updating rules), we used a restricted version for our system. In addition, the following analysis is accurate solely for linear output functions. However, since our architecture works well with either linear or non-linear output functions, the conclusions are valid.

We can rewrite the relevant updating equations of the linear neural network (in matrix form), see ([38] equations 15-19):

$$\psi^{t+1} = Q \cdot J^t \cdot (r^t)^T \qquad (12)$$

$$\Delta J^t = \varepsilon^t \left( \psi^t \left( r^t \right)^T - \Phi \left( \psi^t \cdot \left( \psi^t \right)^T \right) J^t \right) \tag{13}$$

In our case we set

$$Q = I, \ \Phi = diag$$

Consider the following assumptions

[1] The input sequence $r^t$ consists of independent identically distributed, bounded random variables with zero-mean.

[2] $\{\varepsilon^t\}$ is a positive number sequence satisfying: $\sum_t \varepsilon^t = \infty, \ \sum_t (\varepsilon^t)^2 < \infty$. A typical suitable sequence is $\varepsilon^t = \frac{1}{t}, t = 1,2 \ldots$ .

For long times, we denote

$$E\left[ \psi^t \left( r^t \right)^T \right] \to E\left[ J \cdot r \cdot r^T \right] = E[J] \cdot E\left[ r \cdot r^T \right] = J\Sigma \tag{14}$$

$$\lim_{t \to \infty} E\left[ \psi \psi^T \right] = E[J] \cdot E\left[ rr^T \right] \cdot E\left[ J^T \right] = J\Sigma J^T \tag{15}$$

The penultimate equalities in these equations used the fact that the weights converge with probability one to their average value, resulting from the solution of the ODEs. Following [38], we can analyze equations (12)-(13) under the above assumptions, via their asymptotically equivalent associated ODEs

$$\frac{dJ}{dt} = J\Sigma - diag\left( J\Sigma J \right) J \tag{16}$$

with equilibria at

$$J\Sigma = diag(J\Sigma J)J \qquad (17)$$

Since equation (17) is non-linear, an exact (analytic) solution for a given covariance matrix Σ is difficult to obtain. Instead, we solved it numerically by exploiting the same covariance matrix and an initially random weights $J$. In line with our previous findings, we found that constraining $J$ to be non-negative (by a simple cut-off rule) resulted in a hexagonal shape (in the projection of $J$ onto the place cells space; Fig 14). In contrast, when the weights were not constrained they converged to squared-like results.

## Acknowledgments

We would like to thank Alexander Mathis and Omri Barak for comments on the manuscript, and Gilad Tocker, Daniel Soudry and Matan Sela for helpful discussions and advice. The research was supported by the ISRAEL SCIENCE FOUNDATION grants 955/13 and 1882/13, by a Rappaport Institute grant, and by the Allen and Jewel Prince Center for Neurodegenerative Disorders of the Brain. D.D. is a David and Inez Myers Career Advancement Chair in Life Sciences fellow.

# List of references

1. Burgess N (2014) The 2014 Nobel Prize in Physiology or Medicine: A Spatial Model for Cognitive Neuroscience. Neuron 84: 1120-1125.
2. Morris R (2015) The mantle of the heavens: Reflections on the 2014 nobel prize for medicine or physiology. Hippocampus.
3. Eichenbaum H (2015) Perspectives on 2014 Nobel Prize. Hippocampus.
4. O'Keefe J, Dostrovsky J (1971) The hippocampus as a spatial map. Preliminary evidence from unit activity in the freely-moving rat. Brain Res 34: 171-175.
5. O'Keefe J, Nadel L (1978) The hippocampus as a cognitive map. Oxford; New York: Clarendon Press; Oxford University Press. xiv, 570 p. p.
6. Hafting T, Fyhn M, Molden S, Moser MB, Moser EI (2005) Microstructure of a spatial map in the entorhinal cortex. Nature 436: 801-806.
7. Boccara CN, Sargolini F, Thoresen VH, Solstad T, Witter MP, et al. (2010) Grid cells in pre- and parasubiculum. Nat Neurosci 13: 987-994.
8. Sargolini F, Fyhn M, Hafting T, McNaughton BL, Witter MP, et al. (2006) Conjunctive representation of position, direction, and velocity in entorhinal cortex. Science 312: 758-762.
9. Solstad T, Boccara C, Kropff E, Moser MB, Moser EI (2008) Representation of Geometric Borders in the Entorhinal Cortex. Science 322: 1865-1868.
10. Savelli F, Yoganarasimha D, Knierim JJ (2008) Influence of boundary removal on the spatial representations of the medial entorhinal cortex. Hippocampus 18: 1270-1282.
11. Derdikman D, Knierim JJ (2014) Space, Time and Memory in the Hippocampal Formation: Springer.
12. Zilli EA (2012) Models of grid cell spatial firing published 2005-2011. Frontiers in Neural Circuits 6.
13. Giocomo Lisa M, Moser M-B, Moser Edvard I (2011) Computational Models of Grid Cells. Neuron 71: 589-603.
14. Mittelstaedt ML, Mittelstaedt H (1980) Homing by path integration in a mammal. Naturwissenschaften 67: 566-567.
15. Kropff E, Treves A (2008) The emergence of grid cells: Intelligent design or just adaptation? Hippocampus 18: 1256-1269.
16. Si B, Treves A (2013) A model for the differentiation between grid and conjunctive units in medial entorhinal cortex. Hippocampus 23: 1410-1424.
17. Witter MP, Amaral DG (2004) Hippocampal Formation. In: Paxinos G, editor. The Rat Nervous System. 3rd ed. San Diego, CA: Elsevier Academic Press. pp. 635-704.
18. Bonnevie T, Dunn B, Fyhn M, Hafting T, Derdikman D, et al. (2013) Grid cells require excitatory drive from the hippocampus. Nature Neuroscience 16: 309-317.
19. Langston RF, Ainge JA, Couey JJ, Canto CB, Bjerknes TL, et al. (2010) Development of the Spatial Representation System in the Rat. Science 328: 1576-1580.
20. Wills TJ, Cacucci F, Burgess N, O'Keefe J (2010) Development of the Hippocampal Cognitive Map in Preweanling Rats. Science 328: 1573-1576.

21. Oja E (1982) Simplified neuron model as a principal component analyzer. Journal of mathematical biology 15: 267-273.
22. Dai H, Geary Z, Kadanoff LP (2009) Asymptotics of eigenvalues and eigenvectors of Toeplitz matrices. Journal of Statistical Mechanics: Theory and Experiment 2009: P05012.
23. Gray RM (2006) Toeplitz and circulant matrices: A review: now publishers Inc.
24. Montanari A, Richard E (2014) Non-negative principal component analysis: Message passing algorithms and sharp asymptotics. arXiv preprint arXiv:14064775.
25. Kushner HJ, Clark DS (1978) Stochastic Approximation Methods for Constrained and Unconstrained Systems (Applied Mathematical Sciences, Vol. 26): Springer.
26. Wiesel TN, Hubel DH (1963) Effects of visual deprivation on morphology and physiology of cells in the cat's lateral geniculate body. J Neurophysiol 26: 6.
27. Enroth-Cugell C, Robson JG (1966) The contrast sensitivity of retinal ganglion cells of the cat. The Journal of physiology 187: 517-552.
28. Derdikman D, Hildesheim R, Ahissar E, Arieli A, Grinvald A (2003) Imaging spatiotemporal dynamics of surround inhibition in the barrels somatosensory cortex. J Neurosci 23: 3100-3105.
29. Barry C, Hayman R, Burgess N, Jeffery KJ (2007) Experience-dependent rescaling of entorhinal grids. Nature Neuroscience 10: 682-684.
30. Stensola H, Stensola T, Solstad T, Frøland K, Moser M-B, et al. (2012) The entorhinal grid map is discretized. Nature 492: 72-78.
31. Hornik K, Kuan C-M (1992) Convergence analysis of local feature extraction algorithms. Neural Networks 5: 229-240.
32. Freund TF, Buzsaki G (1996) Interneurons of the hippocampus. Hippocampus 6: 347-470.
33. Zass R, Shashua A. Nonnegative sparse PCA; 2006. pp. 1561-1568.
34. Stensola T, Stensola H, Moser M-B, Moser EI (2015) Shearing-induced asymmetry in entorhinal grid cells. Nature 518: 207-212.
35. Mathis A, Stemmler MB, Herz AV (2014) Probable nature of higher-dimensional symmetries underlying mammalian grid-cell activity patterns. arXiv preprint arXiv:14112136.
36. Wei X-X, Prentice J, Balasubramanian V (2013) The sense of place: grid cells in the brain and the transcendental number e. arXiv preprint arXiv:13040031.
37. Sanger TD (1989) Optimal unsupervised learning in a single-layer linear feedforward neural network. Neural Networks 2: 459-473.
38. Weingessel A, Hornik K (2000) Local PCA algorithms. Neural Networks, IEEE Transactions on 11: 1242-1250.

# Appendix

## Movement schema of agent and environment data

The relevant simulation data used:

| Size of arena 10X10 | Place cells width: 0.75 | Place cells distribution: uniform |
|---|---|---|
| Velocity: 0.25 (linear), 0.1-6.3 (angular) | # Place cells: 625 | Learning rate: 1/(t+1e5) |

The agent was moved around the virtual environment according to:

$$D^{t+1} = \text{modulo}(D^t + \omega \cdot Z, 2\pi)$$
$$x^{t+1} = x^t + v \cdot \cos(D^{t+1})$$
$$y^{t+1} = y^t + v \cdot \sin(D^{t+1})$$

where $D^t$ is the current direction angle, $\omega$ is the angular velocity, $Z \sim N(0,1)$ where $N$ is the standard normal Gaussian distribution, $v$ is the linear velocity, and $(x^t, y^t)$ is the current position of the agent.

## Positive-negative disks

Positive-negative disks are used with the following activity rules:

$$\left(r_j^t(x^t, y^t)\right) = \begin{cases} 1 & , \quad \sqrt{(x^t - c_{x,j})^2 + (y^t - c_{y,j})^2} < \rho_1 \\ -\dfrac{\rho_1^2}{\rho_2^2 - \rho_1^2} & , \quad \rho_1 < \sqrt{(x^t - c_{x,j})^2 + (y^t - c_{y,j})^2} < \rho_2 \\ 0 & , \quad \rho_2 < \sqrt{(x^t - c_{x,j})^2 + (y^t - c_{y,j})^2} \end{cases}$$

Where $c_x, c_y$ are the centers of the disks, and $\rho_1, \rho_2$ are the radii of the inner circle and the outer ring, respectively. The constant value in the negative ring was chosen to yield zero integral over the disk.